\documentclass[runningheads,citeauthoryear]{apinv}
\usepackage{epsfig,cite,graphics}
\usepackage[T2A]{fontenc}
\usepackage[cp1251]{inputenc}
\usepackage{longtable}

\begin{document}

\title{Photometric and spectroscopic study of the new FUor star V2493 Cyg}
\titlerunning{Photometric and spectroscopic study of the new FUor star V2493 Cyg}
\author{Evgeni H. Semkov\inst{1}, Stoyanka P. Peneva\inst{1}, Sunay I. Ibryamov\inst{1,2}}
\authorrunning{E. Semkov et al.}
\tocauthor{Evgeni Semkov}
\institute{Institute of Astronomy and National Astronomical Observatory, Bulgarian Academy of Sciences, Sofia, Bulgaria
         \and Department of Theoretical and Applied Physics, University of Shumen, Shumen, Bulgaria
        \newline
        \email{esemkov@astro.bas.bg}
}
\papertype{Research report. Accepted on xx.xx.xxxx}
\maketitle

\begin{abstract}
The recent results from photometric and spectroscopic study of the FUor star V2493 Cyg (HBC 722) are presented in the paper. 
The outburst of V2493 Cyg was registered during the summer of 2010 before the brightness of the star to reach the maximum value.
V2493 Cyg is the first FUor object, whose outburst was observed from its very beginning in all spectral ranges.
The recent photometric data show that the star keeps its maximum brightness during the period September 2013 - Ocober 2016 and the recorded amplitude of the outburst is $\Delta$$V$=5.1 mag.
Consequently, the outburst of V2493 Cyg lasts for more than six years. 
Our spectral observations showed strong variability in the profiles and intensity of emission lines especially for H$\alpha$ line.
We expect that the interest in this object will increase in the coming years and the results will help to explore the nature of young stars.
\end{abstract}

\keywords{Stars: pre-main sequence, Stars: variables: T Tauri, FU Orionis, Stars: individual: V2493 Cyg}

\section{Introduction}

One of the most remarkable phenomenon in the early stages of stellar evolution is the FU Ori (FUor) outbursts.
The prototype of FUors is the eruptive star FU Orionis, located in the Orion star forming region. 
The star was brightened by 6 mag. in 1936 and for a long time was the only one object of its kind. 
FUors are defined as a class of young variables by Herbig (1977) after the discovery of two new FUor objects: V 1057 Cyg and V 1515 Cyg. 
The main characteristics of FUors are an increase in optical brightness of about 4-5 mag., a F-G supergiant spectrum with broad blue-shifted Balmer lines, strong infrared excess, connection with reflection nebulae, and location in star-forming regions (Reipurth \& Aspin 2010, Audard et al. 2014). 
Typical spectroscopic properties of FUors include a gradual change in the spectrum from earlier to later spectral type from the blue to the infrared, a strong Li I ($\lambda$ 6707) line, P Cygni profiles of H$\alpha$ and Na I ($\lambda$ 5890/5896) lines, and the presence of CO bands in the near infrared spectra (Herbig 1977, Bastian \& Mundt 1985). 

The light curves of FUors are varying from one object to another, but in general the rise goes faster than decline in brightness.
A typical outburst of FUor objects can last for several decades or even a century.
Registration of an outburst in the optical wavelengths is considered to be a necessary condition a single pre-main sequence star (PMS) to be accepted as FUor.
Therefore, any new announcement for registration of an eruption from PMS star is welcomed with a great interest by researchers.

FUor stars seem to be related to the low-mass PMS objects (T Tauri stars), which have massive circumstellar disks. 
The widespread explanation of the FUor phenomenon is a sizable increase in accretion rate from the circumstellar disc onto the stellar surface.
The cause of increased accretion is still being discussed.
But the possible triggering mechanisms of FUor outburst could be: thermal or gravitational instability in the circumstellar disk (Hartmann \& Kenyon 1996) and the interactions of the circumstellar disk with a giant planet or nearby stellar companion on an eccentric orbit (Lodato \& Clarke 2004, Pfalzner 2008).
During the outburst, accretion rates raze from $\sim$10$^{-7}$$M_{\sun}$$/$yr to $\sim$10$^{-4}$$M_{\sun}$$/$yr which changes significantly the circumstellar environment. 
The surface temperature of the disk becomes 6000-8000 $K$ and it radiates most of its energy in the optical wavelengths.
For the period of $\sim$100 years the circumstellar disk adds $\sim$10$^{-2}M_{\sun}$ onto the central star and it ejects $\sim$10\% of the accreting material in a high velocity stellar wind. 
Some FUor objects were found to exhibit periodic spectroscopic (Herbig et al. 2003, Powell et al. 2013) or low amplitude photometric (Kenyon et al. 2000, Green et al. 2013, Siwak et al. 2013) variability in short time-scale (days).

During the summer of 2010 a large amplitude outburst from a PMS star located in the dark clouds (so-called "Gulf of Mexico") between NGC 7000 and IC 5070 was discovered (Semkov \& Peneva 2010a, Miller et al. 2011).
The star received designation V2493 Cyg according to the General catalog of variable stars (it is also known as HBC 722, LkH$\alpha$ 188-G4 and PTF10qpf). 
The outburst of V2493 Cyg generated considerable interest and was studied across a wide spectral range.
Simultaneously with the optical outburst appearance of a reflection nebula around V2493 Cyg was observed.
During the FUor type outburst the brightness of the star has increased by about hundred times and this makes visible a part of the diffuse matter around the star.
Such reflection nebulae are typical of all classical FUor objects.
Follow-up photometric observations by Semkov \& Peneva (2010b, 2011), Miller et al. (2011), Leoni et al. (2010), Semkov et al. (2010, 2012, 2014), Lorenzetti et al. (2011), K{\'o}sp{\'a}l et al. (2011, 2016) and Baek et al. (2015) recorded an ongoing light increase in both the optical and infrared.
Follow-up high and low resolution spectroscopic observations by Munari et al. (2010), Miller et al. (2011), Lee et al. (2011, 2015), Lorenzetti et al. (2012) and Semkov et al. (2010, 2012) showed significant changes in both the profiles and intensity of the spectral lines.
The spectral energy distribution (SED) of V2493 Cyg before and during the outburst is discussed in the papers of Miller et al. (2011), K{\'o}sp{\'a}l et al. (2011, 2016) and Gramajo et al. (2014). 
The authors concluded that before the eruption V2493 Cyg was a Class II young stellar object (YSO) according to the evolutionary sequence suggested by Adams et al. (1987).
The Class II YSOs are most often associated with Classical T Tauri stars, in contrast to the Class II that are younger deeply embedded sources, seen only in the infrared region.
The upper limit of the mass of the circumstellar disc is defined by Dunham et al. (2012) and K{\'o}sp{\'a}l et al. (2016).
Perhaps the disk surrounding V2493 Cyg has a relatively low mass (0.01-0.02 $M_{\sun}$) for an object of FUor type.

\section{Observations}

The present paper is a continuation of our photometric and spectroscopic study of V2493 Cyg during the outburst (Semkov et al. 2010, 2012, 201).
We present recent $BVRI$ photometric data of the star in the period September 2013 - October 2016, and new data from spectral observations.
The CCD photometric observations of V2493 Cyg were performed with the 2-m RCC, the 50/70-cm Schmidt, and the 60-cm Cassegrain telescopes of the National Astronomical Observatory Rozhen (Bulgaria) and the 1.3-m RC telescope of the Skinakas Observatory of the Institute of Astronomy, University of Crete (Greece)\footnote{Skinakas Observatory is a collaborative project of the University of Crete, the Foundation for Research and Technology - Hellas,
and the Max-Planck-Institut f\"{u}r Extraterrestrische Physik.} of the Institute of Astronomy, University of Crete (Greece).
Observations were performed with four types of the CCD camera Vers Array 1300B at the 2-m RCC telescope, ANDOR DZ436-BV at the 1.3-m RC telescope, FLI PL16803 at the 50/70-cm Schmidt telescope and FLI PL09000 at the 60-cm Cassegrain telescope.

All the data were analyzed using the same aperture, which was chosen as 4\arcsec in radius (while the background annulus was from 13\arcsec to 19\arcsec) in order to minimize the light from the surrounding nebula and avoid contamination from adjacent stars. 
As references, we used the $BVRI$ comparison sequence of fifteen stars in the field around V2493 Cyg published in Semkov et al. (2010). 
In this way we provided a maximum consistency of the photometric results obtained during the various stages of the photometric observations.
The results of our photometric CCD observations of V2493 Cyg are summarized in Table 1.  
The columns provide the date and Julian date (JD) of observation, $\it IRVB $ magnitudes of V2493 Cyg, the telescope and CCD camera used. 
The typical instrumental errors in the reported magnitudes are $0.01$ for $I$ and $R$-band data, $0.01$-$0.02$ for $V$, and $0.01-0.03$ for $B$-band.

\begin{longtable}{llllllll}
\caption{Photometric CCD observations of V2493 Cyg}\\
\hline\hline
\noalign{\smallskip}  
Date \hspace{1.5cm} &	J.D. (24...) \hspace{2mm}	&	I	\hspace{8mm} & R \hspace{8mm} & V \hspace{8mm} & B \hspace{8mm} & Telescope \hspace{1mm} & CCD	\\
\noalign{\smallskip}  
\hline
\endfirsthead
\caption{continued.}\\
\hline\hline
\noalign{\smallskip}  
Date \hspace{1.5cm} &	J.D. (24...) \hspace{2mm}	&	I	\hspace{8mm} & R \hspace{8mm} & V \hspace{8mm} & B \hspace{8mm} & Telescope \hspace{1mm} & CCD	\\
\noalign{\smallskip}  
\hline
\noalign{\smallskip}  
\endhead
\hline
\endfoot
\noalign{\smallskip}
2013 Sep 04 & 56540.275 & 11.20 & 12.33 & 13.37 & 14.91 &Sch  &FLI\\
2013 Sep 05 & 56541.323 & 11.19 & 12.32 & 13.36 & 14.89 &Sch  &FLI\\
2013 Sep 06 & 56542.382 & 11.19 & 12.35 & 13.39 & 14.92 &Sch  &FLI\\
2013 Sep 07 & 56543.396 & 11.24 & 12.37 & 13.41 & 14.97 &2m   &VA\\
2013 Sep 08 & 56544.264 & 11.30 & 12.47 & 13.50 & 15.04 &2m   &VA\\
2013 Sep 11 & 56547.384 & 11.20 & 12.33 & 13.38 & 14.85 &60cm &FLI\\
2013 Sep 14 & 56550.364 & 11.22 & 12.36 & 13.42 & 14.96 &60cm &FLI\\
2013 Sep 17 & 56553.284 & 11.23 & 12.39 & 13.44 & 14.97 &1.3m &ANDOR\\
2013 Oct 11 & 56577.307 & 11.27 & 12.41 & 13.44 & 14.93 &60cm &FLI\\
2013 Oct 12 & 56578.332 & 11.28 & 12.43 & 13.48 & 14.98 &60cm &FLI\\
2013 Nov 07 & 56604.265 & 11.31 & 12.46 & 13.54 & 15.05 &60cm &FLI\\
2013 Dec 09 & 56636.186 & 11.22 & 12.37 & 13.40 & 14.89 &2m   &VA\\
2013 Dec 28 & 56655.201 & 11.22 & 12.39 & 13.44 & 14.96 &Sch  &FLI\\
2013 Dec 29 & 56656.181 & 11.28 & 12.44 & 13.49 & 15.03 &Sch  &FLI\\
2013 Dec 30 & 56657.194 & 11.27 & 12.40 & 13.47 & 14.98 &Sch  &FLI\\
2014 Jan 23 & 56681.189 & 11.31 & 12.45 & 13.50 & 15.03 &Sch  &FLI\\
2014 Feb 05 & 56694.195 & 11.22 & 12.34 & 13.41 & 14.92 &2m   &VA\\
2014 Mar 22 & 56738.591 & 11.21 & 12.38 & 13.42 & 14.96 &Sch  &FLI\\
2014 May 21 & 56799.426 & 11.24 & 12.38 & 13.42 & 14.97 &Sch  &FLI\\
2014 May 23 & 56801.437 & 11.30 & 12.42 & 13.48 & 15.03 &2m   &VA\\
2014 Jun 23 & 56832.393 & 11.30 & 12.43 & 13.49 & 15.04 &2m   &VA\\
2014 Jun 25 & 56834.357 & 11.29 & 12.44 & 13.50 & 15.03 &2m   &VA\\
2014 Jun 26 & 56835.463 & 11.31 & 12.46 & 13.50 & 15.06 &2m   &VA\\
2014 Jun 28 & 56837.418 & 11.26 & 12.40 & 13.44 & 14.97 &Sch  &FLI\\
2014 Jun 29 & 56838.400 & 11.28 & 12.44 & 13.48 & 15.02 &Sch  &FLI\\
2014 Jul 20 & 56859.398 & 11.28 & 12.43 & 13.50 & 15.01 &60cm &FLI\\
2014 Jul 21 & 56860.392 & 11.23 & 12.39 & 13.44 & 14.96 &60cm &FLI\\
2014 Jul 24 & 56863.340 & 11.29 & 12.47 & 13.50 & 15.03 &Sch  &FLI\\
2014 Jul 25 & 56864.358 & 11.26 & 12.44 & $-$   & $-$   &Sch  &FLI\\
2014 Aug 03 & 56873.295 & 11.26 & 12.44 & 13.52 & 15.09 &2m   &VA\\
2014 Aug 03 & 56873.387 & 11.26 & 12.44 & 13.50 & 15.05 &Sch  &FLI\\
2014 Aug 04 & 56874.409 & 11.14 & 12.25 & 13.30 & 14.78 &Sch  &FLI\\
2014 Aug 18 & 56888.451 & 11.14 & 12.28 & 13.33 & 14.87 &Sch  &FLI\\
2014 Aug 19 & 56889.367 & 11.19 & 12.34 & 13.41 & 14.95 &Sch  &FLI\\
2014 Aug 29 & 56899.303 & 11.20 & 12.37 & 13.45 & 14.99 &1.3m &ANDOR\\
2014 Nov 26 & 56988.183 & 11.19 & 12.35 & 13.38 & 14.93 &Sch  &FLI\\
2014 Nov 28 & 56990.180 & 11.28 & 12.42 & 13.50 & 14.98 &Sch  &FLI\\
2014 Dec 13 & 57005.185 & 11.17 & 12.33 & 13.32 & 14.91 &Sch  &FLI\\
2014 Dec 14 & 57006.222 & 11.19 & 12.36 & 13.33 & $-$   &Sch  &FLI\\
2014 Dec 24 & 57016.175 & 11.20 & 12.38 & 13.42 & 14.99 &2m   &VA\\
2014 Dec 25 & 57017.179 & 11.22 & 12.39 & 13.43 & 15.00 &2m   &VA\\
2015 Feb 21 & 57074.643 & 11.17 & 12.33 & 13.35 & 14.92 &Sch  &FLI\\
2015 Apr 23 & 57136.494 & 11.20 & 12.40 & 13.42 & 14.98 &Sch  &FLI\\
2015 Apr 25 & 57138.494 & 11.22 & 12.39 & 13.44 & 15.00 &Sch  &FLI\\
2015 May 18 & 57161.426 & 11.24 & 12.39 & 13.46 & 14.98 &Sch  &FLI\\
2015 May 19 & 57162.468 & 11.19 & 12.34 & 13.39 & 14.94 &Sch  &FLI\\
2015 May 21 & 57164.461 & 11.19 & 12.34 & 13.37 & 14.94 &Sch  &FLI\\
2015 May 24 & 57167.411 & 11.16 & 12.27 & 13.40 & 14.95 &2m   &VA\\
2015 Jun 12 & 57186.446 & 11.13 & 12.29 & 13.30 & 14.88 &Sch  &FLI\\
2015 Jun 13 & 57187.479 & 11.16 & 12.30 & 13.38 & 14.96 &2m   &VA\\
2015 Jun 16 & 57190.368 & 11.19 & 12.32 & 13.38 & 14.99 &2m   &VA\\
2015 Jul 16 & 57220.364 & 11.13 & 12.30 & 13.34 & 14.89 &Sch  &FLI\\
2015 Jul 17 & 57221.418 & 11.18 & 12.33 & 13.38 & 14.94 &Sch  &FLI\\
2015 Jul 19 & 57223.404 & 11.21 & 12.38 & 13.45 & 15.02 &2m   &VA\\
2015 Jul 20 & 57224.423 & 11.20 & 12.34 & 13.44 & 14.99 &2m   &VA\\
2015 Aug 11 & 57246.356 & 11.17 & 12.31 & 13.37 & 14.92 &1.3m &ANDOR\\
2015 Aug 12 & 57247.394 & 11.18 & 12.35 & 13.42 & 14.97 &1.3m &ANDOR\\
2015 Aug 17 & 57252.385 & 11.17 & 12.31 & 13.36 & 14.94 &2m   &VA\\
2015 Aug 24 & 57259.338 & 11.20 & 12.37 & 13.38 & 14.95 &Sch  &FLI\\
2015 Aug 25 & 57260.343 & 11.20 & 12.37 & 13.39 & 14.98 &Sch  &FLI\\
2015 Sep 03 & 57269.330 & 11.20 & 12.36 & 13.36 & 14.95 &Sch  &FLI\\
2015 Sep 04 & 57270.331 & 11.22 & 12.37 & 13.40 & 15.01 &2m   &VA\\
2015 Sep 05 & 57271.324 & 11.27 & 12.44 & 13.50 & 15.07 &2m   &VA\\
2015 Sep 06 & 57272.294 & 11.28 & 12.43 & 13.50 & 15.07 &2m   &VA\\
2015 Nov 03 & 57330.192 & 11.27 & 12.45 & 13.49 & 15.04 &Sch  &FLI\\
2015 Nov 04 & 57331.232 & 11.28 & 12.46 & 13.52 & 15.06 &Sch  &FLI\\
2015 Nov 05 & 57332.244 & 11.27 & 12.44 & 13.50 & 15.05 &Sch  &FLI\\
2015 Nov 06 & 57333.229 & 11.29 & 12.46 & 13.51 & 15.06 &Sch  &FLI\\
2015 Nov 07 & 57334.221 & 11.27 & 12.45 & 13.50 & 15.03 &Sch  &FLI\\
2015 Dec 12 & 57369.214 & 11.24 & 12.42 & 13.49 & 15.10 &2m   &VA\\
2015 Dec 13 & 57370.188 & 11.32 & 12.49 & 13.58 & 15.17 &2m   &VA\\
2015 Dec 14 & 57371.201 & 11.32 & 12.52 & 13.55 & 15.17 &2m   &VA\\
2015 Dec 15 & 57372.218 & 11.24 & 12.40 & 13.45 & 15.01 &Sch  &FLI\\
2015 Dec 17 & 57374.270 & 11.23 & 12.42 & 13.45 & 15.02 &Sch  &FLI\\
2016 Jan 02 & 57390.196 & 11.21 & 12.40 & 13.37 & 15.02 &Sch  &FLI\\
2016 Feb 06 & 57425.204 & 11.22 & 12.37 & 13.41 & 14.96 &Sch  &FLI\\
2016 Feb 07 & 57426.200 & 11.25 & 12.40 & 13.49 & 15.06 &Sch  &FLI\\
2016 Apr 05 & 57483.520 & 11.22 & 12.37 & 13.44 & 15.06 &2m   &VA\\
2016 Apr 06 & 57484.560 & 11.21 & 12.34 & 13.43 & 15.09 &2m   &VA\\
2016 Apr 07 & 57485.524 & 11.19 & 12.36 & 13.40 & 14.95 &Sch  &FLI\\
2016 Apr 27 & 57506.464 & 11.19 & 12.36 & 13.40 & 14.93 &Sch  &FLI\\
2016 May 13 & 57522.427 & 11.21 & 12.38 & 13.42 & 14.97 &Sch  &FLI\\
2016 May 14 & 57523.428 & 11.24 & 12.42 & 13.48 & 15.03 &Sch  &FLI\\
2016 May 31 & 57540.422 & 11.30 & 12.38 & 13.58 & 14.99 & 2m  &VA\\
2016 Jun 25 & 57565.422 & 11.19 & 12.36 & 13.42 & 14.96 & Sch & FLI\\
2016 Jul 02 & 57572.343 & 11.25 & 12.42 & 13.48 & 14.99 & 2m & VA\\
2016 Jul 11 & 57581.384 & 11.23 & 12.41 & 13.46 & 14.98 & Sch & FLI\\
2016 Jul 12 & 57582.419 & 11.17 & 12.35 & 13.39 & 14.94 & Sch & FLI\\
2016 Jul 13 & 57583.398 & 11.18 & 12.36 & 13.40 & 14.97 & Sch & FLI\\
2016 Aug 01 & 57602.383 & 11.20 & 12.39 & 13.45 & 15.05 & 2m & VA\\
2016 Aug 02 & 57603.367 & 11.19 & 12.39 & 13.43 & 15.02 & 2m & VA\\
2016 Aug 04 & 57605.391 & 11.11 & 12.31 & 13.37 & 14.92 & Sch & FLI\\
2016 Aug 05 & 57606.381 & 11.10 & 12.26 & 13.31 & 14.85 & Sch & FLI\\
2016 Aug 06 & 57607.375 & 11.13 & 12.28 & 13.33 & 14.89 & Sch & FLI\\
2016 Sep 11 & 57643.309 & 11.31 & 12.50 & 13.55 & 15.10 & Sch & FLI\\
2016 Oct 02 & 57664.477 & 11.32 & 12.50 & 13.57 & 15.12 & Sch & FLI\\
\end{longtable}

During the period 2011-2015 a total of five medium-resolution optical spectra of V2493 Cyg were obtained.
The spectral observations are performed in Skinakas Observatory with the focal reducer of the 1.3 m RC telescope and ISA 608 spectral CCD camera ($2000\times800$ pixels, 15$\times15$ $\mu$m).
Two grating 1300 lines/mm and 600 lines/mm and a 160$\mu$m slit were used.
The combination of gratings and slit yield a resolving power $\lambda/\Delta\lambda$ $\sim$ 1500 at H$\alpha$ line for the 1300 lines/mm grating and $\lambda/\Delta\lambda$ $\sim$ 1300 for the 600 lines/mm grating.
The exposures of the objects were followed immediately by an exposure of a FeHeNeAr comparison lamp and exposure of a spectrophotometric standard star.
All data reduction was performed within IRAF. 
The log of spectral observations is summarized in Table 2.

\begin{table}[htb]
  \begin{center}
  \caption{Log of spectral observations of V2493 Cyg with the 1.3-m telescope}
  \begin{tabular}{lcccc}
\hline\hline
\noalign{\smallskip}  
Date \hspace{1cm} & Grating \hspace{6cm}& Slit \hspace{6cm}& Total Exp. [s] \hspace{4cm}& Spectral int. [\AA]\\
\noalign{\smallskip}  
\hline
\noalign{\smallskip} 
2011 Aug 15 & 1300 & 160 & 3600 & 5200$-$7200\\ 
2012 Aug 31 & 2400 & 160 & 3600 & 5750$-$6720\\ 
2012 Sep 01 & 1300 & 160 & 2400 & 5500$-$7500\\
2012 Sep 23 & 1300 & 160 & 3600 & 5400$-$7400\\
2015 Sep 17 & 1300 & 160 & 3600 & 5200$-$7200\\
\noalign{\smallskip}
\hline
 \end{tabular}
  \label{table1}
  \end{center}
 \end{table}  

\section{Results and discussion}

The $BVRI$ light-curves of V2493 Cyg during the period June 2008 $-$ October 2016 are plotted in Fig. 1.
The filled diamonds represent our CCD observations (Semkov et al. 2010, 2012, 2014 and the present paper),
and the filled circles observations from the 48 inch Samuel Oschin telescope at Palomar Observatory (Miller et al. 2011), 
The photometric observation obtained before the outburst displayed only small amplitude variations in all pass-bands typical of T Tauri stars.  
The observational data indicate that the outburst started sometime before May 2010, and reached the first maximum value in September/October 2010.  
Since October 2010, a slow fading was observed and up to May 2011 the star brightness decreased by 1.4 mag. ($V$). 
In the period from May till October 2011 no significant changes in the brightness of the star are observed, its brightness remains at 3.3 mag. ($V$) above the quiescence level.
Since the autumn of 2011, another light increase occurred and the star became brighter by 1.8 mag. ($V$) until April 2013. 
During the period April 2013 - October 2016 the star keeps its maximum brightness showing a little bit fluctuations around it.

\begin{figure}[!htb]
  \begin{center}
   \centering{\epsfig{file=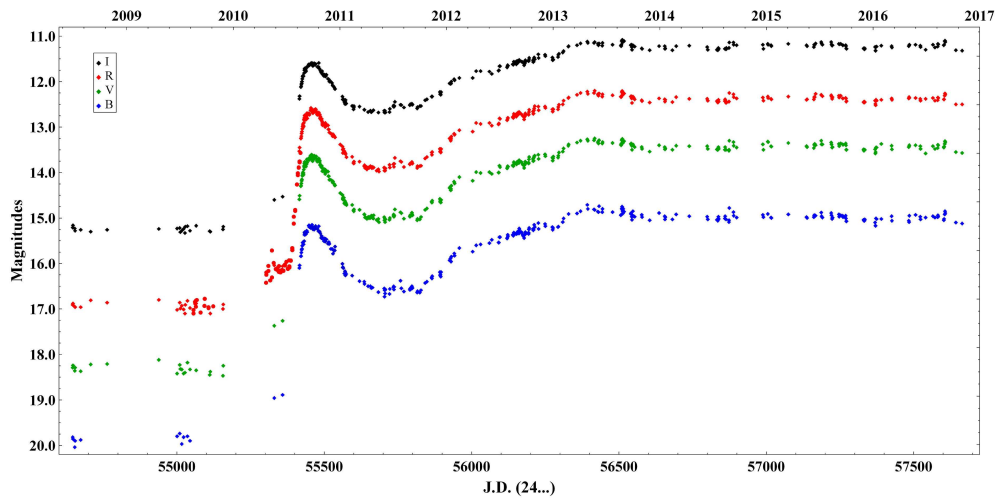, width=\textwidth}}
    \caption[]{$BVRI$ light curves of V2493 Cyg for the period from June 2008 till October 2016}
    \label{countryshape}
  \end{center}
\end{figure}

\begin{figure}[!htb]
  \begin{center}
   \centering{\epsfig{file=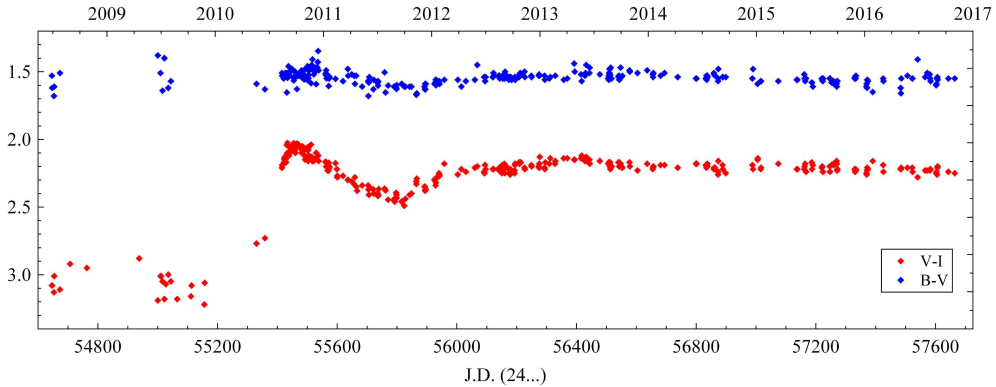, width=\textwidth}}
    \caption[]{Color evolution of V2493 Cyg from June 2008 till October 2016}
    \label{countryshape}
  \end{center}
\end{figure}

Simultaneously with the increase in brightness the star color changes significantly, it become considerably bluer. 
But, while the both indexes $V-I$ and $R-I$ decreased; the $B-V$ index remains relatively constant before and during the outburst (Fig. 2).
Moreover, there is a significant difference in the values of $V-I$ and $R-I$ indexes during the two peaks of brightness (Fig. 3).
During the first local maximum (Aug. 2010 $-$ May 2011) the star is much blue than during the second maximum (Jun. 2011 $-$ Oct. 2016). 
The difference in the $V-I$ indexes at the time of the two peaks in brightness reach 0.2 mag. at the same values of $V$ magnitude.
Such a phenomenon can be explained by a gradual expansion of the emitting region around the star.

\begin{figure}[!htb]
  \begin{center}
   \centering{\epsfig{file=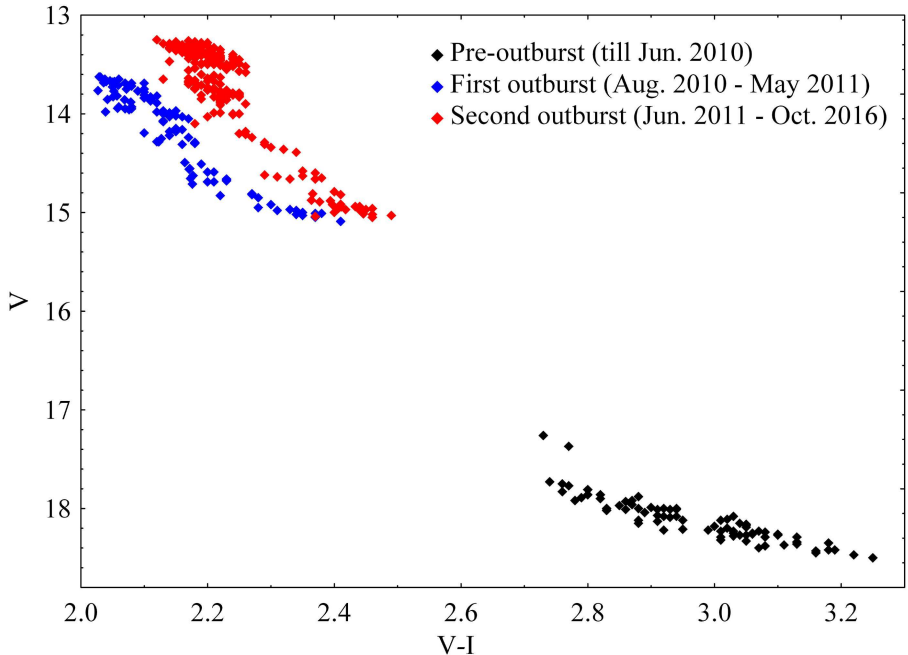}}
    \caption[]{$V/V-I$ diagram from our $VI$ photometric CCD data.}
    \label{countryshape}
  \end{center}
\end{figure}

In order to study the spectral variability of V2493 Cyg during its photometric evolution, we have made efforts to conduct and spectral monitoring.
Our spectral observations during the outburst showed strong variability in intensity and profiles of spectral lines despite of the relatively quiet photometric state.
This phenomenon is well noticeable by comparing the P Cyg profiles of the H$\alpha$ hydrogen line from spectra obtained during the recent years (Fig. 4).
Moreover exceptional night-to-night variability of profiles was registered too.
Such a phenomenon has been observed and for the other stars of this type (Herbig et al. 2003).

\begin{figure}[!htb]
  \begin{center}
   \centering{\epsfig{file=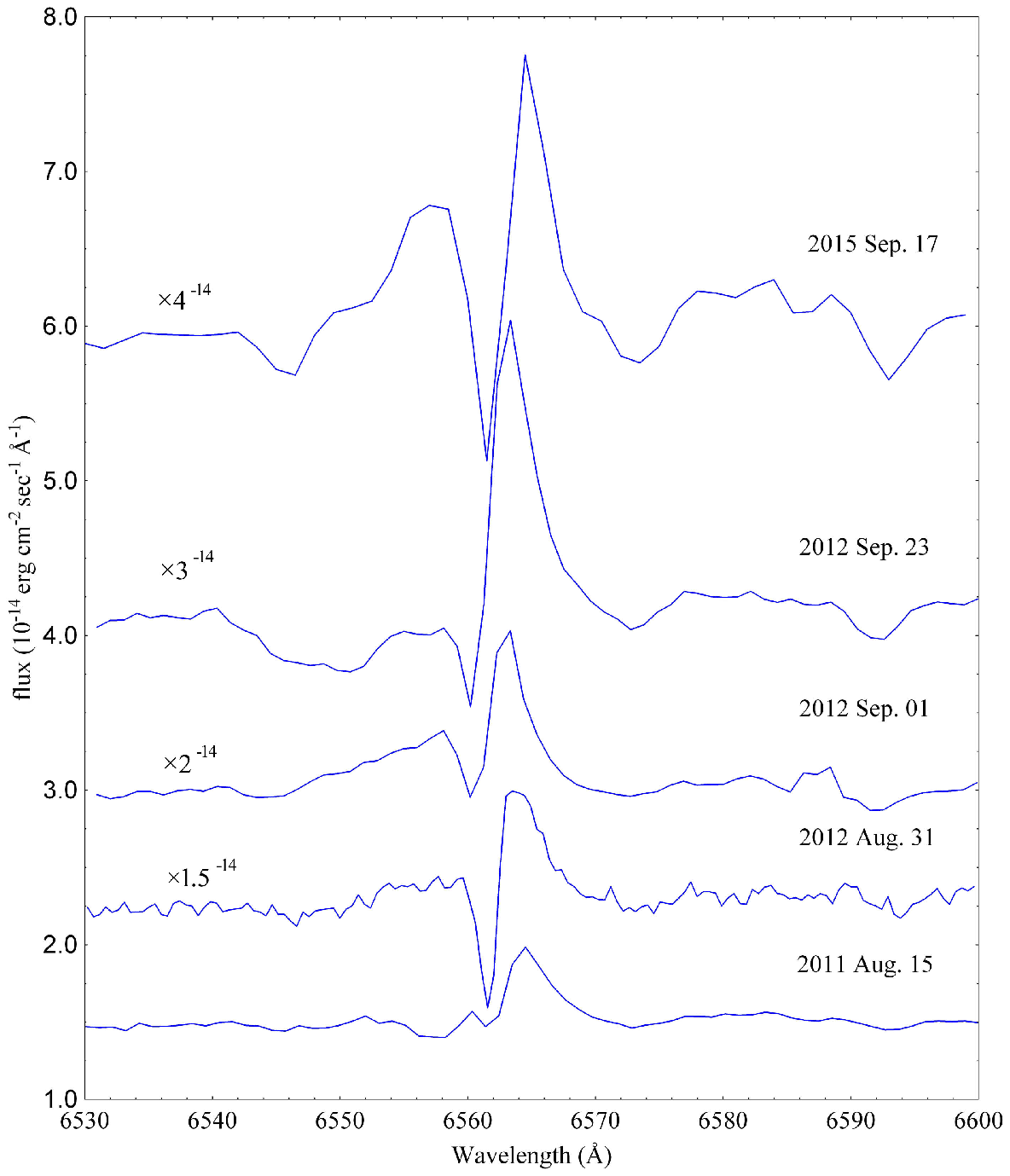}}
    \caption[]{Profiles of H$\alpha$ line during the period of spectral observations. The spectra are shifted in the flux for a better viewing.}
    \label{countryshape}
  \end{center}
\end{figure}

For some years, we have made efforts to construct the historical light curves of several other FUor and FUor-like objects such as: V1735 Cyg (Peneva et al. 2009), Parsamian 21 (Semkov \& Peneva 2010c), V733 Cep (Peneva et al. 2010), V1647 Ori (Semkov \& Peneva 2012), and V582 Aur (Semkov et al. 2013).
To realize this study, we use both data from recent CCD photometric monitoring as well as photometric data from the photographic plate archives.
Our results suggest that each object of FUor type has a characteristic long-term light curve, which distinguishes it from other objects. 
The shape of the observed light curves of FUors may vary considerably in the time of rise, the rate of decrease in brightness, the time spend at maximum brightness, and the light variability during the set in brightness. 
The different kind of light curves is evidence that the reasons causing the outburst and primary physical conditions (mass and temperature of the circumstellar disk) vary at each FUor object.

The light curve of V2493 Cyg from all available photometric observations is also somewhat unique.
The rate of increase in the brightness (the fastest ever recorded) was followed by a very rapid fall in brightness.
During the period of rise in brightness and the first months after the maximum, the light curve of V2394 Cyg is similar to the light curves of the classical FUor object V1057 Cyg and FU Ori itself.
But the most remarkable feature of the light curve of V2493 Cyg is the repeated rise in brightness in 2011-2013 and the reaching of a second maximum in brightness.

\section{Conclusion}
Due to the wide interest in optical and infrared photometric observations V2493 Cyg became the FUor object with the most well-studied light curve at the present.
Regardless of the very small changes in brightness over the last three years, the spectral observations recorded significant variations in the spectral lines.
Despite the very low mass of the circumstellar disk (see Introduction) V2493 Cyg still keeps the maximum state of brightness and photometrically it is no different from the classical FUor stars.
We plan to continue our spectroscopic and photometric monitoring of the star during the years and strongly encourage similar follow-up observations.

{\it Acknowledgments:} The authors thank the Director of Skinakas Observatory Prof. I. Papamastorakis and Prof. I. Papadakis for the award of telescope time. 
This research has made use of the NASA's Astrophysics Data System Abstract Service, the SIMBAD database and the VizieR catalogue access tool, operated at CDS, Strasbourg, France. 
This research was supported partly by funds of the project RD-08-81 of Shumen University.

\end{document}